\documentstyle[prl,aps,epsfig,twocolumn,amsmath]{revtex}        

\renewcommand{\epsilon}{\varepsilon}

\begin{document}

\draft
\twocolumn[\hsize\textwidth\columnwidth\hsize\csname @twocolumnfalse\endcsname
\title{Tailoring  symmetry groups using external alternate fields.}

\author{I. Junier and J. Kurchan}
\address{ 
\it P.M.M.H. Ecole Sup{\'e}rieure de Physique et Chimie Industrielles,
\\
10, rue Vauquelin, 75231 Paris CEDEX 05,  France}

\date{\today}

\maketitle

\begin{abstract}
 Macroscopic systems with continuous symmetries 
subjected to  oscillatory fields  have  phases and
transitions that are qualitatively different from their equilibrium ones.
 Depending on the amplitude and frequency of the fields applied,
Heisenberg ferromagnets  can become XY or
 Ising-like --- or, conversely,  anisotropies can be compensated ---
thus changing the nature of the ordered phase and the topology of  defects.
The phenomenon  can be viewed as a dynamic form of  {\it order by
disorder}.

\end{abstract}

\vskip2pc]

\narrowtext

The effect of alternate fields on interacting many-body systems 
has been a subject of long-standing  interest.
In particular, magnetic hysteresis   has been extensively studied
for  clean or disordered
Ising-like systems \cite{DTH,Ac3,Set1,Set2,THL}.

 A much less studied problem is the effect of a periodic drive on  
 systems having a continuous symmetry either in a pure or a disordered
\cite{Rao1,Rao2,HSODhar,rava} case; 
even though,
 as we shall see, the existence of soft modes 
make the systems respond  in striking and
interesting ways.
Years ago, Rao {\em et al.}
 \cite{Rao1,Rao2} and Dhar and Thomas \cite{HSODhar}
studied   a driven  clean $N$-component Heisenberg ferromagnet, solvable
analytically
 in the large-$N$ limit  for any dimension $d$. 
Due to the fact that at $N=\infty$
 almost any configuration is perpendicular to the field,
 one can not study thus, phenomena that depend on a delicate balance between
a tendency of the system to order longitudinally or transversely to the field.

In this paper we explore the full problem for $d>2$ and $N \geq 2$. We
 firstly treat  the  XY ($N=2$) model and show that, surprisingly enough, there
 are in fact
 {\em three} kinds of ferromagnetic phases,
 with the magnetization longitudinal, transverse 
and canted with respect to the field, respectively.
We show this directly for the mean-field ($d=\infty$) case, 
and extend the result
for all dimensions $d>2$ by use of low frequency and low 
temperature expansions. 
The same analytic 
methods show that for the Heisenberg case ($N\geq 3$) there is always
 transverse order\cite{HSODhar}, again in all $d>2$. 
 In addition we show that  one can tailor in this case the
symmetry group from $O(3)$ to $O(2)$ or to $Z_2$ 
by applying one or more a.c. fields: a possibility we argue
is quite general for systems with continuous symmetries.

Our motivation  is that an understanding of this system
 is a base to explore the connections
with wider class of 
forced systems with continuous symmetries, including liquid crystals 
 \cite{liqcryst}, lamellar polymers \cite{polym},
crystalline defects, ferromagnetic conductors \cite{Holm}, etc.

Consider the $O(n)$ ferromagnet
\begin{equation}
E_{int}= - \frac{1}{2d} \sum_{ij} {\vec{S}}_i \cdot {\vec{S}}_j
\label{oo}
\end{equation}
where
$2d$ is the number of neighbors, itself depending on
lattice dimensionality and topology.
The ${\vec{S}}$ are $N$-dimensional  vectors either of fixed norm {\it (hard
spins)}, or whose length may fluctuate according to a  soft-spin term.
The system is coupled to an ac field:
\begin{equation}
E_{h}= - \cos(\omega t ) \; \sum_{i} {\vec{h}} \cdot {\vec{S}}_i 
\label{ooo}
\end{equation}
and we shall consider the dynamics in the {\em strongly dissipative
  limit}:
\begin{equation}
{\dot{S}}_i^\alpha = - \frac{\partial E}{\partial S_i^\alpha} +
\eta_i^{\alpha}
\label{pp}
\end{equation}
where  $\eta_i^{\alpha}$ is a
white gaussian noise of variance $2T$. The energy $E$ is the sum of (\ref{oo}), (\ref{ooo})
 and a {\it soft-spin} term fixing
the spin length or, alternatively, a Lagrange multiplier for {\it hard} spins.
In (\ref{pp}), we are neglecting precessional effects (which bring in a host
of interesting phenomena \cite{Rao3}).

\vspace{.3cm}

{\bf Selection: order by disorder.}

\vspace{.1cm}

Consider the `hard' spin  model at zero temperature, subjected to an a.c.
 field. 
 We shall use the example of the XY model,
but the argument is valid for any number of components and in any
number of dimensions.
Using polar coordinates,
we have for the angle $\theta_i$:
\begin{equation}
{\dot{\theta}}_i= -\frac{1}{d} \sum_j \sin(\theta_i-\theta_j) - h
  \cos(\omega t) \cos \theta_i
\end{equation}
This always admits a solution in which all spins move in phase
$\theta_i=\theta, \;\;\;\forall i$ (inhomogeneous 
solutions  decay into this one)
which makes interaction terms zero and we have for any geometry:
\begin{equation}
\theta(t) =
2 \tan^{-1} \left(e^{-[\frac{h}{\omega} \sin(\omega t) + k]}\right)- 
  \frac{\pi}{2}
\label{cf}
\end{equation}
By considering all possible values of the integration constant $k$, we
conclude that at zero temperature solutions are possible in which the
total magnetization vector oscillates
around {\em any possible angle}, and  without hysteresis.
 Unlike the case without field, these solutions are
no longer related by a continuous symmetry, and only the discrete
symmetries
$\theta \rightarrow -\theta$ and $\theta \rightarrow \pi-\theta$ remain.
(In the case of $N>2$,  a  symmetry subgroup survives: for example, in
$N=3$ the spins evolve with
 $\theta$ as in (\ref{cf}) and with  constant $\varphi$;
solutions with the same $k$ but different $\varphi$ are related by an
$O(2)$ symmetry.)

In situations like this one, when one has  at $T=0$ solutions
 that are  unrelated by a symmetry,  thermal or quantal fluctuations
 will
generically
select a subset of them  (unless of course they destroy the order altogether).
This  phenomenon has been named {\it order by disorder} \cite{OBD},
 as it is the fluctuations
that are responsible for a reduction in the number of solutions.
In practice, if we prepare the system at small $T$ in one of the possible 
$T=0$
 solutions, it will  drift  to the selected average
 angle: there is a secular perturbation 
 acting on
timescales much slower than the vibration frequency. Temperature also brings in  
hysteresis and hence dissipation.

\vspace{.3cm}

{\bf  Paramagnetic, transverse, longitudinal and canted solutions.}

\vspace{.1cm}

In an $O(N)$ ferromagnet, one can have four types of 
 phases,
which can be best distinguished by considering the magnetization in
the direction parallel and perpendicular to the field $M_h(t)$ and  
$M_\perp(t)$,
  the angle $\theta(t)=\tan^{-1}(M_\perp(t)/M_h(t))$, and the
 corresponding averaged quantities: ${\overline{M^\alpha}}\equiv \frac{1}{\tau}
\oint dt M^\alpha(t) $ and  $\bar \theta \equiv \frac{1}{\tau}
\oint dt \; 
\theta(t)$
\begin{itemize}
\item {\em Paramagnetic}: The magnetization follows the field with a 
delay (hysteresis): 
$M_\perp(t)=0$ and $ \overline{M_h} = 0$.
\item {\em Longitudinal $(\theta(t)=0 \; \; or \; \;\theta(t)=\pi)$}:
 the magnetization points  in the 
direction of the field: $M_\perp(t)=0$, but $\overline{M_h} \neq 0$.
\item {\em Transverse $(\bar \theta=\pi/2$ or $\bar \theta=-\pi/2)$:} 
the magnetization  has a non-zero component $M_\perp(t)\neq 0$ 
orthogonal to the field,
the component parallel to the field has zero time-average $\overline{M_h}=0$.
\item {\em Canted $(0<\bar \theta <\pi/2$ or $\pi/2<\bar \theta <\pi)$:} 
The magnetization evolves around an oblique angle with the field's direction:
$\overline{M_\perp} \neq 0$ and $ \overline{M_h}  \neq 0$. 
\end{itemize}
The longitudinal solution is doubly degenerate. In the XY case, the canted 
solutions have degeneracy four and the transverse ones two. Both of them 
are continuously degenerate
for $N>2$ (one solution per plane determined by $(M_h,M_\perp)$).

Dynamical phase transitions to a magnetized
phase are well attested
 in the case of an Ising system \cite{lo,Ac1,Ac2}.
For continuous  $O(N=\infty)$ systems, Rao {\it et al} (\cite{Rao1,Rao2})  
found  a 
dynamical phase transition on increasing frequency 
from a paramagnetic state with $\overline{M}=0$
to a ferromagnetic regime. 
Subsequently,
 Dhar and Thomas \cite{HSODhar} pointed out that the order is in fact
always transverse, and   
$\overline{M_h} = 0$ \cite{Ovshini}.
They also studied the case $N\geq 2$ 
in finite dimensions within the unmagnetized phase \cite{HISDhar,SOCDhar}.

\vspace{.3cm}

{\bf Dynamic mean-field approximation.}

\vspace{.1cm}

The dynamic mean-field approximation has the advantage of being
completely solvable, and one can easily get a complete phase diagram.
The equations consist of a single-spin equation and
a self-consistency condition:
\begin{eqnarray}
{\dot{S}}^\alpha &=&  (M^\alpha + h^\alpha)(t) -
 \lambda(t) S^\alpha +\eta^{\alpha} \nonumber\\
M^\alpha(t) &=& \langle S^\alpha(t) \rangle_{single\;spin}
\label{ii}
\end{eqnarray}
where $\lambda$ is a Lagrange multiplier imposing the spin length.
One can solve the system (\ref{ii}) by considering the evolution of
the set of expectation values. For example, for the 
 XY model, using the Fokker-Plank equation associated with (\ref{ii}) 
 we write an exact infinite system of equations for 
$x_n \equiv \langle \cos(n \theta) \rangle$, 
$y_n \equiv \langle \sin(n \theta)
\rangle$, $n=1,...$: 
\begin{eqnarray}
{\dot{y}}_n&=&-n^2Ty_n+\frac{1}{2}nx_1(y_{n-1}-y_{n+1})
+m (x_{n-1}+x_{n+1})\nonumber\\
{\dot{x}}_n&=&-n^2Tx_n+\frac{1}{2}nx_1(x_{n-1}-x_{n+1})
-m (y_{n-1}+y_{n+1})
\nonumber\\
{\dot{y}}_1&=&(\frac{1}{2}-T)y_1-\frac{1}{2}x_1y_2-\frac{1}{2}(y_1+h(t))x_2+\frac{1}{2}h(t)
\nonumber\\
{\dot{x}}_1&=&(\frac{1}{2}-T)x_1-\frac{1}{2}x_1x_2-\frac{1}{2}(y_1+h(t))y_2
\label{equa}
\end{eqnarray}
with $m(t)=n(y_1+h(t))/2$. 
For the Heisenberg case one has to study the 
dynamics of the expectation values of the spherical harmonics $Y_{lm}$.

We have numerically solved the system (\ref{equa}),
keeping as many modes as necessary,  
for various values of $h,\omega, T$. 
The main results are summarized in the figures.

\vspace{0.3cm}

\begin {figure}
\begin{center}
\input {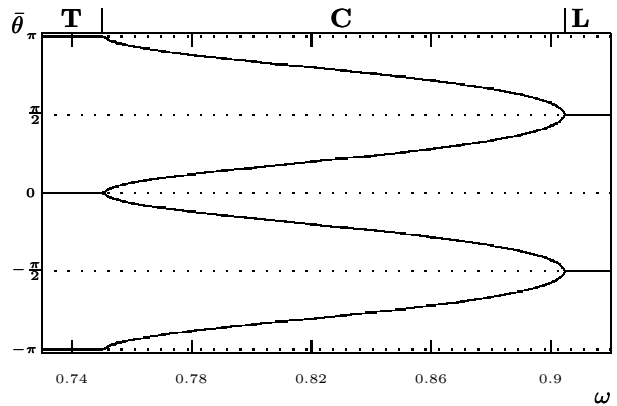}
\vspace{.2cm}
\caption{
$\bar \theta$ as a function of $\omega$ for the XY mean
  field 
model, $h=1$, $T=0.2$. The transitions are second order.}
\end{center}
\end {figure}

FiG. 1. shows  the value of $\bar \theta$ vs
$\omega$ 
at temperature $T=0.2$ and $h=1$. We see second order transitions, as
frequency decreases,
from longitudinal (L) to canted (C), and from canted to transverse (T).
FIG. 2. gives the general phase diagram of the system at $h=1$ (the dashed line
at T=0.5 represents the para-ferro transition at zero field). 
In the limit of high frequencies $\omega \rightarrow \infty$, $h/\omega \rightarrow 0$,
there is a  dynamical critical temperature $T_d \sim .42$ from a longitudinal to a
 transverse phase.

\begin{figure}
\begin{center}
\psfig{file=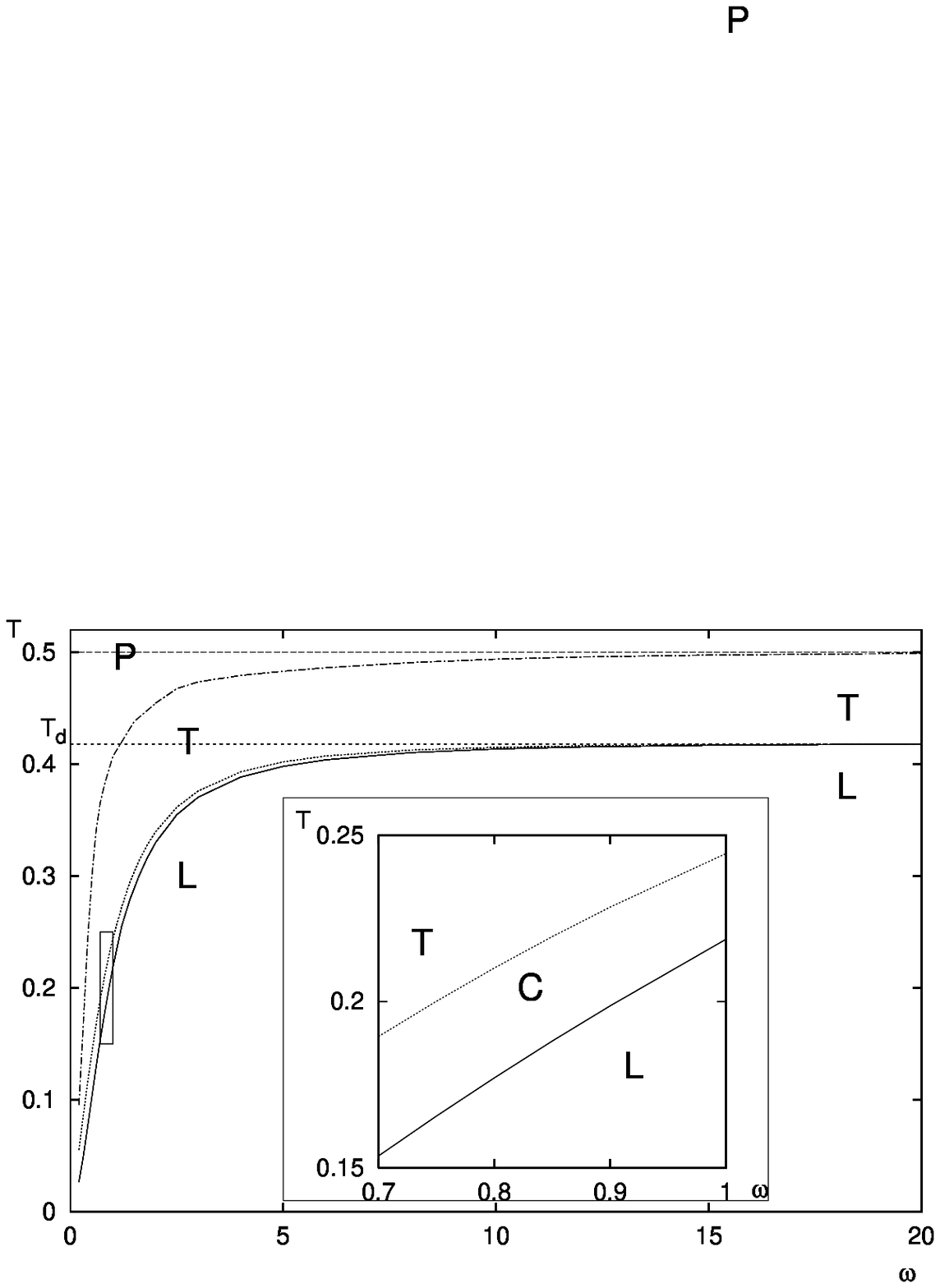,width=8.5cm,height=5cm}
\vspace{.2cm}
\caption{
$T$-$\omega$ Phase diagram for the XY mean field model, $h=1$. }
\label{lstar}
\end{center}
\end{figure}

In the Heisenberg case, we have only found paramagnetic or
 transverse behavior. 
Below we confirm   this by showing  that, in all dimensions $d \geq 3$,
no longitudinal order
appears in
the first order in low-temperature  expansion  
(the most favourable case, as seen in Fig. 2),
   thus supporting the 
conjecture of \cite{HSODhar} that
the Heisenberg model, and {\em a foriori} all the $N>2$
 cases,  
 are  in this sense qualitatively different from 
the XY model.
However, as we shall see below, {\em one can still select a single direction} 
in a Heisenberg case, 
but with the application of {\em two} fields. 

Let us stress here that at very low frequencies mean-field is potentially very misleading,
as the physics may be dominated by nucleation effects which this
approximation neglects.

\vspace{.3cm}

{\bf Weak fields and low frequencies:  a general mechanism for transverse selection.}

\vspace{.1cm}

 In the limit of  low frequencies and weak fields 
($\omega \rightarrow 0$,
 $h/\omega$  constant), a general simple argument can be given
to show that there is transverse selection for any $N \geq 3$ and $d \geq 3$:
Consider a field applied along the $z$ axis, and split it
 in  components aligned $h_M(t)$ and normal $h_\perp(t)$
to the instantaneous magnetization. Because by assumption the field is weak and 
varies slowly,
the effect of the aligned component of the field is to change the magnetization norm $M$
 linearly and adiabatically:
$M(t)=M_o + C  h_M(t)$, where $M_o$ 
is the unperturbed norm and $C$ a positive susceptibility.
On the other hand, the effect of the perpendicular field is to modify the angle $\Omega$ between
 the magnetization vector and the equatorial plane as: $ J(M) \dot \Omega = h_\perp(t)$.
The mobility coefficient $ J^{-1}(M)$ depends only upon the norm  $M$ due to symmetry.
Putting this together we get:
\begin{equation}
J(M) \dot \Omega =h(t)\cos\Omega \;\;\; ; \;\;\; 
M=M_o + C h(t)  \sin\Omega
\label{Trans}
\end{equation}
and using the weak field approximation:
\begin{equation}
 \dot \Omega =\frac{h(t) \cos\Omega}{J(M_o)+ C J'(M_o) h(t)  \sin\Omega} 
\label{Trans1}
\end{equation}
It is easy to check that this equation 
yields transverse selection provided $J'(M_o)>0$, i.e. when
the  mobility decreases with the magnetization. 
Let us note that an inhomogeneous nucleation mechanism would need a critical droplet size
of $h^{-1/2}$, and a free-energy barrier $\sim h^{-(d-2)/2}$, which in the present
regime requires much longer times compared to $\omega^{-1}$. 

We can now explain in words the selection as a ratchet mechanism:
 suppose the magnetization starts at an
angle in the first quadrant. During the semicycle when the magnetic field points upwards 
it simultaneously  rotates
 the magnetization towards the $z$ axis and it stretches the  norm.
 In the negative cycle, the rotation is towards the $xy$ plane and the norm is 
shortened. The slight changes in the norm make the positive cycle less efficient
(if $J'(M_o)>0$) 
  than the negative cycles, and hence the vector has a net drift 
 towards the transverse direction on each cycle.
The same mechanism applies for {\em soft spins} in all dimensions at zero temperature.

 We have checked that indeed $J'(M_o)$ is positive in all cases. 
Near the critical point   $J(M)\propto M$,
implying that the selection mechanism dissappears at the paramagnetic transition. 
 
\vspace{.3cm}

{\bf Low $T$ expansion: longitudinal selection}

\vspace{.1cm}

These models are obviously exactly solvable at $T=0$,
and  an expansion in powers of $T$ is feasible  {\em in any
  dimension}. In the XY case, one decomposes
 $\theta_i(t)=\langle\theta_i(t)\rangle+\tilde{\theta}_i$ (averages over the
 thermal noise), 
and one writes an evolution equation
for the expectation values $\langle\theta_i\rangle(t)$, 
$\langle\tilde{\theta}_i\tilde{\theta}_j\rangle(t)$, 
$\langle\tilde{\theta}_i\tilde{\theta}_j \tilde{\theta}_k\rangle(t), ...$.
To leading order in $T$ only two-point correlations are 
necessary. In the most general case, we have:
\begin{eqnarray}
\frac{1}{2}\frac{d}{dt}\langle\tilde{\theta}_a\tilde{\theta}_b\rangle&=&  T \delta_{ab}
-\sum_j A_{aj}\langle\tilde{\theta}_b(\tilde{\theta}_a-\tilde{\theta}_j)\rangle
-h \langle\tilde{\theta}_a\tilde{\theta}_b\rangle \sin \langle\theta_a\rangle
\nonumber \\
\frac{d \langle\theta_a\rangle}{dt}&=&h(t) 
(1-\frac{1}{2}\langle{\tilde{\theta}_a}^2\rangle)\cos \langle\theta_a\rangle
\label{D1}
\end{eqnarray}
where $A_{ai} \propto \frac{1}{d}$ if $a,i$ are neighbors in $d$-dimensional space 
and zero otherwise. 
This system is linear in the correlations. One can assume translational
 invariance with one-point functions  independent of the site and
two-point functions depending only on the distance between sites.
The system then becomes exactly solvable
in Fourier basis (details in \cite{Additional}). For hard spin systems, it always yields   
{\em  longitudinal} selection 
for any $d \geq 3$. Thus, 
we have shown the existence of a longitudinal and a transverse phase 
 for the XY-model
 in finite dimension $d \geq 3$. The existence of the intermediate canted phase
 we have obtained analytically within 
mean-field (large $d$)  has been checked with simulations.

 The generalization to the Heisenberg ($N=3$) case  of the low-temperature expansion
is immediate, one has to  consider two-point correlations of both   
 angles $\theta_i$ and $\phi_i$ \cite{Additional}. One can  show that 
 there is no selection to order  $T$ 
in any dimension $d \geq 3$. Since the lower temperatures are the most
favourable for a longitudinal order, this is evidence that   there
is no longitudinal phase at all.

\vspace{.3cm}

{\bf High frequencies}

\vspace{.1cm}

Another instructive method is the high-frequency expansion in which the 
external field
can be treated perturbatively. We  start from a
system
in equilibrium under an infinitesimal field in a direction $\xi$. In
 terms of the linear $\chi_{\alpha \beta}^\xi$ and higher susceptibilities,
the variation of the magnetization vector $M_\alpha$
 in presence of the a.c. field $h_\alpha(t)$ is given to second order as: 
\begin{equation}
\int dt' \chi_{\alpha \beta}^\xi(t,t')h_\beta(t')+
\int dt' dt''
\chi_{\alpha \beta \gamma}^\xi(t,t',t'')h_\beta(t')h_\gamma(t'') 
\nonumber
\end{equation}

The linear term does not contribute to the drift if $\oint
h_\alpha(t)dt=0$. The quadratic term does and, in the large $\omega$ limit,
the contribution reads: 
\begin{equation}
\frac{\Delta \overline{M^\alpha}}{\tau} =\frac{1}{\omega^2}\sum_{\beta \gamma}
v_{\alpha \beta \gamma}
{\overline{h_\beta h_\gamma}}  \;\; ; \;\;
 v_{\alpha \beta \gamma} \equiv \lim_{\stackrel{t-t' \rightarrow
     \infty}{t''\rightarrow t'}
}  \frac{\partial
  \chi_{\alpha \beta \gamma}^\xi(t,t',t'')}{\partial t'} 
\label{two}
\end{equation}
where $\overline{h_\beta h_\gamma}\equiv \oint
h_\beta(t) h_\gamma(t)dt \neq 0$. Specialising to the mean-field case, 
already the first order in $T$ and in $\omega^{-2}$
gives for the XY case a longitudinal selection corresponding to
 an effective potential
$\sim Th^2\omega^{-2} \sin(2\theta)$.

Now, if the field is a sum of two components 
acting at right angles with a ratio of frequencies larger than two, 
interferences 
appear to the third order and their 
contribution is additive up to  $\omega^{-2}$.  We have checked 
in the hard spin case that  two 
orthogonal fields 
at two high enough frequencies $\omega$ and $3\omega$, 
make the  Heisenberg system Ising-like:
 the magnetization is transverse to
one field and longitudinal to the other. 
The same effect can be obtained with a field with 
constant modulus that rotates
on a plane.

\vspace{.3cm}

{\bf Perspectives}

\vspace{.1cm}

One can forsee several applications of these ideas in different
fields. Changing the symmetry group of a system
leads to a change in  the  topology of the defects: thus an alternate
 field may perhaps be used as a tool  to study in detail the role these play
in the transitions \cite{Holm}, in the phase-ordering kinetics, in
 the anomalous Hall effect (since it involves the interaction
of electrons with the topological defects) \cite{Holm}, etc.
Similarly, the {\it chirality scenario} \cite{Kawa} for the spin-glass
transition (which assumes that spin glasses are essentially
isotropic) could be put to test experimentally by monitoring the
effect of an increased and a decreased effective anisotropy.
Subjecting ordered liquid crystals to a.c. fields might be a way to
directly observe what becomes of the order and the textures, a strategy for which
  there 
are interesting precedents \cite{yyy}.


 Such richness of amusing and perhaps useful
 phenomena is in contrast with the extreme poverty of 
  concepts for these out of equilibrium problems, since there are few
  qualitative ideas to guide us before having an actual solution, as we have for
  equilibrium thermodynamic systems.

\end{document}